\documentclass[11pt]{article}
\usepackage[letterpaper,margin=1in]{geometry}
\usepackage[T1]{fontenc}
\usepackage{lmodern}
\usepackage[protrusion=true,expansion=true]{microtype}
\usepackage{amsmath,amssymb,amsthm}
\usepackage{xcolor}
\usepackage{aliascnt}

\definecolor{darkred}{RGB}{139,0,0}

\DeclareRobustCommand{\Lemma}{\textcolor{darkred}{Lemma}}

\newtheoremstyle{darkredplain}
  {\topsep}{\topsep}
  {\itshape}
  {}
  {\color{darkred}\bfseries}
  {.}
  {.5em}
  {}

\usepackage[
  colorlinks=true,
  linkcolor=darkred,
  citecolor=darkred,
  urlcolor=darkred
]{hyperref}

\theoremstyle{darkredplain}
\newtheorem{theorem}{Theorem}
\newaliascnt{lemma}{theorem}
\newtheorem{lemma}[lemma]{\Lemma}
\aliascntresetthe{lemma}

\DeclareRobustCommand{\refprefix}[1]{#1~}
\labelformat{theorem}{\refprefix{Theorem}#1}
\labelformat{lemma}{\refprefix{Lemma}#1}
\labelformat{section}{\refprefix{Section}#1}
\labelformat{subsection}{\refprefix{Subsection}#1}

\newcommand{\Local}{\mathsf{Local}}
\newcommand{\ECC}{\mathsf{ECC}}
\newcommand{\Fin}{\mathsf{Fin}}
\newcommand{\shift}{\operatorname{shift}}
\newcommand{\Forall}{\forall}
\def\lcp{\textsf{lcp}}

\title{\textls[15]{Explicit} $O(1)$ \textls[15]{Tree Code}}
\author{Samuel Dittmer\thanks{Stealth Software Technologies,
  Inc. Email: {\em sam@stealthsoftwareinc.com} \; \;
  https://orcid.org/0000-0003-0018-6354} \ \ \ \ \ \  \ \ \ \ \  Rafail
  Ostrovsky\thanks{UCLA, Email: {\em rafail@cs.ucla.edu}\; \;
  https://orcid.org/0000-0002-1501-1330}}
\date{}

\begin{document}

\maketitle

\begin{abstract}
We construct an efficiently computable constant rate and constant relative
distance infinite tree code.
\end{abstract}

\section{Introduction}
The existence of an infinite binary tree code with a constant
relative distance, a constant rate, and a constant alphabet size
was proven in the seminal work of Schulman over thirty years
ago~\cite{Schulman92,Schulman1993,Schulman1996}. A tree code reads one input
bit at a time and outputs a symbol from the alphabet $\Sigma$ in a streaming
fashion. Its rate is $1/\log_2|\Sigma|$. Assume that two input streams
first differ at time $a$. The binary tree code requires that for every
$t \ge a$,  the output must differ in at least $\delta(t-a+1)$ positions
starting from $a$ until time  $t$, even if the two input streams agree at
every position after $a$. Schulman introduced tree codes for interactive
communication in the early 90's and proved that a fixed output alphabet
and constant relative distance $\delta>0$ suffice for streams of arbitrary
length~\cite{Schulman92,Schulman1993,Schulman1996}. How can we efficiently
map an input stream of bits to an output stream of symbols while achieving
both guarantees? This paper presents a deterministic online encoder that
achieves a constant rate and constant relative distance. It emits one output
symbol as soon as any input bit arrives and requires no advance bound on
the stream length. In our construction, $t$ input symbols take $O(t)$ word
operations in total. Furthermore, we guarantee $O(\log^2(t+2))$ worst-case
word operations per each incoming input bit, as $t$ grows.

For inputs of length at most $N$, our bound uses words that grow as well,
and are of size $\lceil C\log^2(N+2)\rceil$ bits, where $C$ is a constant.
On these words, we count arithmetic operations as a single step, including
multiplication and division. If counting bits instead of words, an output
symbol depth $t$ can also be computed by our algorithm from its input prefix
in $O(t^2)$ operations on bits.

\paragraph{Prior bounds.}
Prior unconditional polynomial-time constructions (for the parameters in
the table below) either kept the relative distance constant and allowed the
output alphabet to grow, or kept the output alphabet fixed and obtained the
relative distance guarantee that decreases as $n$ grows. The table gives
these bounds for the binary input stream.

\begin{table}[htbp]
\centering
\renewcommand{\arraystretch}{1.15}
\begin{tabular}{|l|c|c|}
\hline
\textbf{Construction} & \textbf{Rate} & \textbf{Relative distance} \\
\hline \hline
Evans--Klugerman--Schulman~\cite{EvansEtAl1994}
  & $\Omega(1/\log n)$ & constant \\
\hline
Cohen--Haeupler--Schulman~\cite{CohenEtAl2018}
  & $\Omega(1/\log\log n)$ & constant \\
\hline
Gelles et al.~\cite{GellesEtAl2016,BenYaacovEtAl2022}
  & constant & $\Omega(1/\log n)$ \\
\hline
Ben Yaacov--Cohen--Yankovitz~\cite{BenYaacovEtAl2022}
  & constant & $\Omega(1/(\log\log n)^2)$ \\
\hline
Yankovitz~\cite{Yankovitz2026MDS}
  & constant & $\Omega(1/\log\log n)$ \\
\hline
{\em This paper} &  constant & constant \\
\hline
\end{tabular}
\end{table}

Braverman showed both constant rate and constant relative distance,
but  with subexponential run-time construction by composing shorter
tree codes~\cite{Braverman2012}. Moore and Schulman gave a poly-time
construction where distance  follows from a conjecture regarding exponential
sums~\cite{MooreSchulman2014}. Ben Yaacov, Cohen, and Narayanan also
proposed another construction whose distance guarantee follows from another
conjecture about minors of Pascal matrix~\cite{BenYaacovEtAl2021}.

A recent rate bound of Cohen, Schulman, and Srivastava assumes a stronger
distance condition, called immediacy~\cite{CohenEtAl2026}. In its simplest
form, immediacy requires a distance guarantee over prescribed window
lengths starting at every position where the inputs differ. Our construction
satisfies the original tree-code guarantee on every window starting at the
very first differing bit. This allows windows starting at later differing
bits to have poor distance, consistent with~\cite{CohenEtAl2026}.

\section{Overview of Our Techniques}

\paragraph{Local encoders}
Our encoder has a local component for short intervals and a packet
component for longer ones. The local component starts with a length-$r$
convolutional encoder, following the original convolution that Schulman
credits to Rabani and the subsequent binary vector formulation in
Peczarski~\cite{Schulman1996,Peczarski2004,Peczarski2006}.

We repeat this convolution idea twice (the second time separately on residue
classes of time modulo $r$). This two-pass trick already appeared in the
interleaved convolutional encoder of Lodge and Young~\cite{LodgeYoung1993}.
Also, Braverman's product gives a distance at depth $r^2$ with $2^{O(r)}$
run time~\cite{Braverman2012}. In our variant, for every $1\le\ell\le r^2$,
at least $\ell/4$ of the $\ell$ local output symbols differ starting from
the first differing input from two possible divergent streams. This idea
protects the stream during short runs.

\paragraph{Packet encoding of a large-alphabet tree code.} Our construction
  is built on top of a large-alphabet tree code
$TC_{\mathbb{F}_p^d}: \left(\mathbb{F}_p^{d}\right)^n \to
\left(\mathbb{F}_p^d \times \mathbb{F}_p^{8d}\right)^n$ with
$d = \Theta(\log p)$ and $n=2^{7d} < p$. Therefore, if $\lambda$ is the
bitlength of the input alphabet, our large-alphabet tree code has depth
$2^{\Theta(\sqrt{\lambda})}$. Previous large-alphabet building blocks in
papers cited in the table below, with constant-rate, constant-distance,
have depth polynomial in the input alphabet bit length (see e.g.
\cite{CohenEtAl2018, Yankovitz2026MDS}), so they employed techniques such
as the Lagged Tree code introduced in \cite{CohenEtAl2018} to interleave
multiple tree codes. In our construction, we can instead break the message
up into blocks of packets. We treat each packet as an input symbol from
the large alphabet tree code, and write the output symbol for that packet
alongside the local encoding of the next block of messages.

\paragraph{Reciprocal sums.} Our large-alphabet tree code consists of
  copying the input symbol and writing an additional check symbol computed
  from all previous packets. The bitlength of the check symbol is a constant
  multiple of the bitlength of the input symbol. The purpose of this check
  symbol, informally, is to force disagreement in the check symbols when
  there are long stretches of agreement in the input streams after the
  initial point of divergence.

At a high level, this aspect of our approach is similar to the tree code
over $\mathbb{Z}$ constructed in \cite{CohenEtAl2018}. In that work, the
authors use a Newton basis for polynomials over $\mathbb{R}$ and show that
the number of non-trivial integer zeros at non-negative integers is bounded
above by the polynomial's sparsity with respect to the Newton basis. Our
approach uses a result over $\mathbb{F}_p$ that shows a reciprocal sum
$$\sum_{j < t} \frac{w_j}{t-j}$$
has fewer zeros than the number of nonzero coefficients $w_j$ (for
a fixed nonempty set of $j$'s as $t$ varies.) We prove this bound
as~\ref{reciprocal}. Similar to~\cite{CohenEtAl2018}, the sparsity bound
alone is insufficient to prove constant relative distance, since placing
the nonzero $w_j$ terms on triangular numbers gives a gap-by-gap distance
guarantee of only $\Theta(1/\sqrt{n})$. To improve this guarantee to
constant distance, we perform a shift operation:

\paragraph{Shifting check terms.} We use packets that lie in a vector of
  field elements in $\mathbb{F}_p^{d}$ instead of the field $\mathbb{F}_p$,
  and perform a \textit{shift} operation in a larger vector space
  $\mathbb{F}_p^{8d}$ on all terms in the checksum of the form $w_j/(t-j)$.
  As a result, only some terms in the sum contribute to the leftmost
  potentially nonzero entry of the vector, modifying our analysis of the
  sparsity of the reciprocal sum.

Yankovitz also analyzed sums of shifted vectors, and also looked
at the first coordinate where the sums could differ. He then looked
at rows that tie at that coordinate, and where these rows pass each
other~\cite{Yankovitz2025Shifts,Yankovitz2026MDS}. Our shift is much
shorter and is bounded by the address length, which allows us to take
$d = \Theta(\log p)$. Our shift is based on a measurement of proximity (the
longest common prefix), which grows coarser the farther apart the terms are.
Therefore, packets are more likely to be tied the further back in the past
history they lie and they can remain tied for many consecutive time steps.
Finally, packets get the same shift once the current address grows beyond
their common-prefix block. We define our longest common prefix operator and
show its key properties in~\ref{lcp}.

\paragraph{Distances across packets.}

The main combinatorial step of our proof is to establish contiguous
intervals on which the same set of earlier differing packets attains
the first possible coordinate. We show that such intervals cover at
least $1/8$ of the packet times past the first difference in the two
streams. We show that these bounds give relative packet distance at least
$1/24$. (We do not attempt to optimize any constant.) The rest of the
construction is standard. We ECC-encode completed packet labels and transmit
them during the next packet arrival together with the local code that
protects short distances. This is similar to the construction of Evans,
Klugerman, and Schulman~\cite{EvansEtAl1994} as well as the constructions
in~\cite{Braverman2012,Pudlak2016}. We run finite encoders with overlapping
horizons to get an unbounded tree code, similar to~\cite{BenYaacovEtAl2022}.

For more efficient encoding, we amortize the computation of the check
vectors by scheduling convolutions for pairs of blocks on which the shift is
constant, similar to the ideas of~\cite{FischerStockmeyer1974}.

\paragraph{Definitions and our result.}
All logarithms are base 2. For integers $u<v$, the notation $[u,v)$ denotes
the integer positions $u,u+1,\ldots,v-1$.

An \emph{online encoder} reads input bits one bit at a time. A positive
integer $T$ denotes the current input position, and $x_1,\ldots,x_T$ denotes
the bits that the encoder has read so far. After reading these bits, it
emits a label $E(x_1,\ldots,x_T)$ from a fixed finite alphabet $\Sigma$.

The encoder is a \emph{tree code of relative distance $\delta$} if the
following  holds for every input length $T$ and every pair of unequal inputs
$x_1,\ldots,x_T$ and $y_1,\ldots,y_T$. Let $a$ be their first position
where the two sequences diverge. The tree-code guarantee is that among
$a,\ldots,T$, at least a $\delta$ fraction of labels is different:
$$
  \bigl|\{t\in[a,T+1):E(x_1,\ldots,x_t)\ne E(y_1,\ldots,y_t)\}\bigr|
  \ge \delta(T-a+1)
$$

When an encoder is defined for depth at most $D$ we call it a \emph{finite
encoder of horizon $D$}. For a finite-horizon encoder, the tree-code
distance is required only for input lengths $\leq D$.

We state our results in two ways: as bit operations or word operations,
both on the RAM model. We use a RAM machine with a word size that is
at least $\lceil C\log^2(N+2)\rceil$ bits and a depth of at most $N$.
$C$ is a sufficiently large constant. In the word-operation model, we
count arithmetic on words, including multiplication,  division, boolean
operations, shifts, and indexing memory, as unit cost. The bit-operation
model counts bit operations instead. $N$ fixes the word size for these
bounds but does not change our (streaming) encoder.

\begin{theorem}\label{main}
There is a uniform tree code with an alphabet of size $O(1)$ and constant
relative distance. A label at depth $t$ can be computed in $O(t)$ word
operations or $O(t^2)$ bit operations. The first $t$ labels can be computed
on-line in $O(t)$ total word operations, with $O(\log^2(t+2))$ worst-case
word operations per input bit.
\end{theorem}

We stress that we do not optimize any of the constants, and there are
multiple ways to make these substantially smaller. Furthermore, the paper
prioritizes clarity of exposition, even at the expense of larger constants.

\section{Preliminaries: Local encoders for short runs}\label{local-section}

We identify a binary string of length $m$ with a vector in $\mathbb F_2^m$,
where $\mathbb F_2$ is the field with two elements. All vector arithmetic in
this section is over $\mathbb F_2$. For $U,V\in\mathbb F_2^m$, $U\oplus V$
denotes the bitwise XOR.

\subsection{Encoder for length \texorpdfstring{$r$}{r}
  run}\label{first-encoder}

We chose a fixed positive integer $r$ (with $r=O(\log(D+2))$,
where $D$ is the depth of the input tree) and seven-bit vectors
$A_0,\ldots,A_{r-1}\in\mathbb F_2^7$. Fix the first vector to be
$$
  A_0=(0,0,0,0,0,0,1)
$$
The last bit of each remaining vector must be zero. Its other six bits are
unrestricted. Set $x_s=0$ for input positions $s\le0$. The ``first-stage''
label $Y_t\in\mathbb F_2^7$ (for runs of length $r$) at position $t$ is
\refstepcounter{equation}\label{eq:short}
$$
  Y_t=x_tA_0\oplus x_{t-1}A_1\oplus\cdots\oplus x_{t-r+1}A_{r-1}
  \eqno(\theequation)
$$
Thus, each vector is included in the XOR exactly when its corresponding
input bit is $1$. When $r$ is $1$, the expression consists only of its
(first) term. Test every length-$r$ binary input beginning with $1$. For
each such input and every prefix length $h=1,\ldots,r$, we require the
condition that more than $\frac h2$ of the labels $Y_1,\ldots,Y_h$ must be
nonzero. We will run a brute force search and choose the lexicographically
first coefficient list passing all these tests.

\begin{lemma}[Short encoder]\label{short}
The search succeeds in $O(2^{7r}r^2)$ word operations. If two inputs first
differ at some position, then for every $1\le h\le r$, more than half of the
first $h$ labels beginning at that position differ between the two inputs.
\end{lemma}

\begin{proof}
Once we prove existence, the string will be found by brute force. Choose
the six free bits of each vector $A_1,\ldots,A_{r-1}$ at random. Consider
a prefix length $h$ and an input prefix beginning with $1$. Its first
label is the fixed nonzero vector $A_0$. For $t \in \{2, \ldots, h\}$,
consider the effect of $x_1$ (which we assume to be $1$)  on $Y_t$.
It  adds $A_{t-1}$ (for $t=1$, this term is $A_0$) to $Y_t$, even when
$A_1,\ldots,A_{t-2}$ are fixed. Thus, the probability that $Y_t$ is zero
is at most $\frac1{64}$ (since there are $6$ free bits). Applying this
repeatedly, for any specified set of $k$ positions among $2,\ldots,h$,
these  labels are zero with probability at most $64^{-k}$. A failed
length-$h$ test requires at least $\lceil\frac h2\rceil$ zero labels.
There are at most $2^h$ choices for a set of positions witnessing this
failure. There are $2^{h-1}$ input prefixes beginning with $1$. Let $F_h$
be the event that at least one input prefix of length $h$ beginning with
$1$ fails its length-$h$ test. Thus the probability of $F_h$ is at most
$\;2^{h-1}\cdot2^h\cdot64^{-\frac{h}{2}}=2^{-h-1}$. Note that the length-$1$
test is always successful. A second union bound gives
$$
  \Pr[\text{some test fails}]\le\sum_{h=2}^{r}2^{-h-1}
$$
The corresponding infinite sum is
$$
  \sum_{h=2}^{\infty}2^{-h-1}=\frac14
$$
Hence some coefficient list passes every test. There are $2^{6(r-1)}$
candidate lists and $2^{r-1}$ test inputs. Computing all labels of one
test input and checking all its prefixes takes $O(r^2)$ word operations.
Therefore, the search has the required bound. Finally, take the bitwise
XOR of two divergent input streams. It is zero before the first differing
position and is $1$ at that position. Since~\eqref{eq:short} commutes with
bitwise XOR, so a nonzero output on this difference input is exactly an
unequal pair of original labels.
\end{proof}

\subsection{Extending to \texorpdfstring{$r^2$}{r\texttwosuperior} run}
  \label{second-encoder}

The short tree code of length $r$ has been found using brute-force search,
but the code of length $r^2$ can be constructed without any additional
search.

For a seven-bit label $Y_t\in\mathbb F_2^7$, $Y_t[k]\in\mathbb F_2$ is the
$k$th bit of $Y_t$. Use the zero label for nonpositive time coordinates
$t$. For each $k$ from $1$ through $7$, define a second-stage label
$V_{t,k}\in\mathbb F_2^7$
$$
  V_{t,k}=Y_t[k]A_0\oplus Y_{t-r}[k]A_1\oplus\cdots\oplus
  Y_{t-(r-1)r}[k]A_{r-1}
$$
With $r$ fixed and the input sequence understood, write $\Local_t\in\mathbb
F_2^{49}$ for the local output at position $t$. It is the following 49-bit
string, formed by concatenating the seven vectors $V_{t,k}\in\mathbb F_2^7$:
$$
  \Local_t=(V_{t,1},\ldots,V_{t,7})
$$
Each residue class of time modulo $r$ has its own seven short encoders. At
each input position, only the seven encoders for the current residue class
receive new bits and produce their seven-bit labels.

\begin{lemma}[Local distance]\label{local}
The local output $\Local_t$ depends only on the last $r^2$ input bits. If
two inputs first differ at $a$, then every comparison of length $\ell\le
r^2$ beginning at $a$ has at least $\frac\ell4$ unequal local outputs.
\end{lemma}
\begin{proof}
The first stage uses delays up to $r-1$. The second stage adds delays up to
$r(r-1)$. Their maximum combined delay is
$$
  (r-1)+r(r-1)=r^2-1
$$
This proves the dependence claim.

For distance, take the bitwise XOR of the two inputs. All its
bits before position $a$ are zero, and its bit at position $a$ is
$1$. The $\ell$ compared positions are written in consecutive rows
$R^{1},R^{2},\ldots,R^{\lceil\frac{\ell}{r}\rceil}$ of length $r$, starting
at position $a$, with the last row being possibly incomplete.

Mark a column if its first-row label $Y_t$ is nonzero. ~\ref{short}
says that more than half of every prefix of the first row is marked.
Consequently, more than half the positions in each complete row belong in
marked columns. The same holds for the incomplete final row. In total, more
than $\frac\ell2$ compared positions are located in marked columns.

Fix one marked column and choose a nonzero coordinate of its first label,
denoted by $Y_t[k]$. Along that column, this coordinate is the input to
one of the short encoders for the second stage. The first input bit of this
second-stage short encoder within the comparison is $1$ and all its earlier
input bits are zero. The column has at most $r$ positions. ~\ref{short}
therefore makes more than half of its second-stage labels nonzero. At each
such position, the full local output is nonzero as well.

The columns are disjoint. Thus, we can sum the contributions of the columns
and get at least $\frac\ell4$ nonzero local outputs on the difference input,
and hence at least that many unequal outputs on the original inputs.
\end{proof}
\section{Technical Lemmas}

We now give two technical lemmas we will need to prove the correctness of
our construction. The first deals with reciprocal sums over $\mathbb{F}_p$.
The second provides key properties of the longest common prefix function
$\lcp_r(i, j)$ defined as the length of the shared prefix of $i$ and $j$
when written as binary substrings of some fixed length $r$.

\subsection{A fact about reciprocal sums - they do not vanish too often}

\begin{lemma}\label{reciprocal}
For a prime $p$ and a nonempty set  $A\subseteq\mathbb F_p$, it holds that
$\Forall j\in A$, fix a nonzero coefficient $w_j\in\mathbb F_p$ and define
$$
  R(t)=\sum_{j\in A}\frac{w_j}{t-j}
$$
where $t\in\mathbb F_p\setminus A$. Then, $R$ has at most $|A|-1$  roots in
$\mathbb F_p\setminus A$.
\end{lemma}

\begin{proof}
When we make a common denominator, the numerator becomes:
$$
  F(z)=\sum_{j\in A}w_j
  \prod_{k\in A\setminus\{j\}}(z-k)
$$
Observe that each summand is a polynomial of degree $|A|-1$. We want to
show that  $F$ is a nonzero polynomial.  Fix any  $a\in A$. Every summand
for  $j\ne a$ contains the factor $z-a$ and therefore becomes zero at $z=a$.
Therefore,
$$
  F(a)=w_a\prod_{k\in A\setminus\{a\}}(a-k)\ne0
$$
since $w_a\ne0$ and $a-k\ne0$ for each $k\in A\setminus\{a\}$. Thus, $F$ is
a nonzero polynomial. It has degree at most $|A|-1$, and hence,  has at most
$|A|-1$ roots. $\forall \; t\in\mathbb F_p\setminus A$,
$$
  R(t)=\frac{F(t)}{\displaystyle\prod_{k\in A}(t-k)}
$$
Since every factor in the denominator is nonzero, the denominator is nonzero
at this $t$. Hence, $R(t)=0$ if and only if $F(t)=0$. This proves the claim
(since this polynomial has at most $|A|-1$ roots).

\end{proof}

\subsection{Observations on longest common prefixes}

Let $0 \leq i < j < 2^r$. Then the longest common prefix $\lcp_r(i, j)$
is the longest shared prefix of $i$ and $j$ when written as $r$-bit binary
strings. Equivalently, it is the largest value $k$ such that
$$\lfloor \frac{i}{2^{r-k}} \rfloor = \lfloor \frac{j}{2^{r-k}} \rfloor.$$
Wherever possible, we omit the term $r$ from $\lcp_r(i, j)$. Since the
values $i, j$ generally correspond to nodes in a binary tree of fixed
depth, the length is generally clear from context. Informally, we view
$\lcp(i, j)$ as a measurement of proximity that grows coarser as $i$ and $j$
grow further apart. It has the following properties that we will verify now
and use later:

\begin{lemma}
\label{lcp}
Let $h = 2^k$ be a power of $2$ and $c$ be a nonnegative integer. For the
addressing to be well-defined, we assume $2(c+1)h\leq 2^r$.
Define $$J = [2ch, 2ch+h)$$ Then:
\begin{enumerate}
  \item For $i < j$, $\lcp(i, j)$ is non-increasing with respect to $j$.
  \item For $i < j < t$, $\lcp(j, t) - \lcp(i, t)$ is non-increasing with
    respect to $t$.
  \item For  fixed $i < 2ch$ and $j \in J$, $\lcp(i, j)$ is constant.
  \item For $i < j$ with $i,j \in J$, $\lcp(i, j) > \lcp(i, 2ch+h)$.
\end{enumerate}
\end{lemma}
\begin{proof}
First, assume that no $r$-bit address greater than $j$ leaves the longest
common prefix $P$ of $i$ and $j$. Then for every $t>j$, $\lcp(i,t)=|P|$.
This proves $(1)$ in that case. In turn, we can deduce $(2)$ by applying
$(1)$ to $\lcp(j,t)$. In all other cases, $k$-boundary  (see below) is
well-defined.
\begin{enumerate}
  \item Let the longest common prefix of $i < j$ be the bitstring $P$, with
    $|P| = \ell$. Then, since $i < j$, $i$ begins with the prefix $P0$ and
    $j$ begins with the prefix $P1$. Let $k$ be the smallest integer greater
    than $j$ with a prefix of length $|P|$ different from $P$, which we call
    the \emph{$k$-boundary of $i$ and $j$}. Then for $j < j_0 < k$ we have
    $\ell = \lcp(i, j) = \lcp(i, j_0) > \lcp(i ,k)$. Even more generally,
    any common prefix of $i$ and $t>j$ is also shared by $j$ since it forms
    an interval. Thus $\lcp(i, j)$ is non-increasing with respect to $j$.
  \item Again, let the longest common prefix of $i < j$ be the bitstring
    $P$ of length $\ell$, with $i$ beginning $P0$ and $j$ beginning
    $P1$, and let $k$ be the smallest integer greater than $j$ with a
    length $|P|$ prefix different from $P$. If $j < t < k$, then $t$ also
    begins with $P1$, so its common prefix with $i$ is $P$ and its common
    prefix with $j$ is at least (and begins with) $P1$. Thus we have
    $\lcp(j, t) - \lcp(i, t) > 0$ and $\lcp(i, t) = \ell$ for $t < k$, and
    since $\lcp(j, t)$ is non-increasing with respect to $t$ by (1), the
    expression $\lcp(j, t) - \lcp(i, t)$ is also non-increasing with respect
    to $t$.

  If $t \geq k$, then $\lcp(j, t) = \lcp(i, t)$, since the first bit where
    $t$ differs from $P$ is its first disagreement with both $i$ and $j$.
    We therefore have $\lcp(j, t) - \lcp(i, t) = 0$ for $t \geq k$. Thus
    $\lcp(j, t) - \lcp(i, t)$ (when $j < k-1$) decreases as $t$ goes from
    $k-1$ to $k$, and then remains constant afterwards.

  \item The first element $2ch$ of $J$ has (at least) $k + 1$ final
    zeros. Let $P$ be the prefix of length $r-k - 1$ of $2ch$. Then the
    interval $J$ is precisely the set of integers with prefix $P0$. Thus
    for $i <2ch$, and $j \in J$, $\lcp_r(i, j) = \lcp_{r-k}( \lfloor i/h
    \rfloor, 2c)$, which does not depend on $j$.
  \item As above, let $P0$ be the leading prefix of $2ch$. Then $i,j$ begin
    with $P0$,but $2ch+h$ begins with $P1$, so $\lcp(i,j) \geq r-k > r-k-1 =
    \lcp(i, 2ch+h)$.
\end{enumerate}
\end{proof}

\section{Packet encoder}
\label{packets}

\subsection{Overview}
\label{overview}

For the packet encoder, we use a fixed power-of-two horizon $D$ (i.e., the
length of the input to the finite-horizon tree-code), measured in input
bits, set as follows. Once we pick $r$ as in \ref{local-section}, (and
setting $r=d$ below) our horizon will be $D=\Theta(r^2 \cdot 2^{7r})$. We
pick parameters as follows. We choose a positive integer $d$, the number
of coordinates in each packet vector, and also a power-of-two packet length
$b$: for each candidate $d$, let $b$ be the largest power of two no greater
than $d^2$. Choose the smallest $d$ for which
$$
  2^{7d}b\ge D
$$
Since $b$ is the largest power of two no greater than $d^2$, we have
$$
  b\le d^2<2b
$$
Equivalently,
$$
  \sqrt b\le d<\sqrt{2b}
$$
Thus $d=\Theta(\sqrt b)$. For horizon $D$, we keep the size of  $d$ and
$b$ fixed. The way we choose these parameters ensures that  $b\le D$, as
verified in ~\ref{finite}. Therefore,  $b$ divides $D$.

For the packet encoder, the input consists of exactly $D/b$ consecutive
packets of $b$ bits each. A \emph{packet position} is the index of the
packet, starting from $0$. Thus packet number  $i$, where $0\le i<D/b$,
contains exactly the input bits at positions $ib+1$ through $(i+1)b$.
Packet contents form a vector $B_i\in\mathbb F_2^b$. The local encoder of
\ref{local-section}, with $r=d$, protects comparisons up to one packet long
because $b\le d^2$. Let $p$ be the least prime greater than $2^{7d}$. To
represent a complete packet $B_i\in\mathbb F_2^b$ over this field, append
zeros to obtain $d^2$ bits, then split these bits into $d$ chunks of $d$
bits each. Their integer values form a packet vector $v_i\in\mathbb F_p^d$.
This padding changes only the representation of a packet. To each completed
packet $i$ we attach a \emph{check vector} as defined in the next few
subsections, and denoted
$$
  q_i\in\mathbb F_p^{8d}
$$
Thus the check vector has exactly $8d$ coordinates, each in $\mathbb F_p$.
It is computed by shifting, scaling, and adding the earlier packets from
$\mathbb F_p^d$, with each shifted vector in $\mathbb F_p^{8d}$. The pair
$(B_i,q_i)$ is the \emph{packet label}. We prove that a constant fraction of
these labels differ after the first differing packet.

\bigskip
In \ref{ecc} we encode each packet label as bits and then apply
standard linear ECC code to the result. This output is then divided into
constant-size chunks that are sent during the next packet transmission, one
chunk per input bit, together with the local output in $\mathbb F_2^{49}$
at that bit time. The local encoder runs continuously across all packet
boundaries.  This is the entire fixed horizon tree code.

For infinite tree codes, we keep two horizons, and switch between them, as
explained in \ref{infinite}.

\subsection{The setup: packets, shifts, and longest common prefix}
\label{packet-rule}

We will use $D,d,b,p$ as explained in the overview, and set
$$
  n=2^{7d}
$$
Observe that packet length is $\Theta(d^2)$ bits. There will be exactly
$D/b$ packets, with $D/b\le n=O(D/b)$. The size parameters estimates are
further analyzed in ~\ref{finite}. For $n$ we can always pick a prime $p$ so
that $n<p<2n$. Therefore, any element of  $\mathbb F_p$ can be represented
using $O(d)$ bits. For tree-code input bits $x=(x_1,\ldots,x_D)\in\mathbb
F_2^D$, packet $i$ for $0\le i<D/b$, has exactly input bits:
$$
  B_i=(x_{ib+1},\ldots,x_{(i+1)b})\in\mathbb F_2^b
$$
For packets, we can view the packet sequence as
$$
  (B_0,\ldots,B_{D/b-1})\in(\mathbb F_2^b)^{D/b}
$$
A packet is not defined at each time step, but rather with frequency $1/b$.
That is, packet $i$ is complete when bit $(i+1)b$ arrives.

To convert a packet to $v_i\in\mathbb F_p^d$, first append zeros to $B_i$
to reach exactly $d^2$ bits. We then split the result into $d$ chunks of
$d$ bits each. For $0\le k<d$, let $v_i[k]\in\mathbb F_p$ is the integer
represented by the $d$-bit chunk $k$. The first bit of each chunk is the
most significant bit. Every chunk treated as an integer has a value that
is less than $2^d<p$, so representing chunks in $\mathbb F_p$ loses no
information.

One key operation in our constructions is shifting by $s$ and inserting data
from a smaller vector into a slightly larger one: $\mathbb F_p^d \rightarrow
\mathbb F_p^{8d}$. Shift $s$ ranges $0\le s<7d$, where shifting an input
vector $v\in\mathbb F_p^d$  is the following operation: we insert $s$ zeros
in the front and pad the result to form a vector:
$$\shift_s(v)\in\mathbb F_p^{8d}$$
For each coordinate $0\le k<8d$, the rule is
$$
  \shift_s(v)[k]=v[k-s]
$$
Here $v[k]$ is zero for $k<0$ or $k\ge d$. No information is lost, as
shifting a vector for some integer $s$ between $0$ and $7d-1$ merely shifts
and moves the $d$ elements of the original vector into a bigger vector.

Another key operation of our construction is to compute {\em longest common
prefix (lcp)} of two packet locations in the stream. For distinct packet
indices $i,j\in\{0,\ldots,D/b-1\}$, write their binary representations with
exactly $7d$ bits,
including leading zeros. Define $$\lcp(i, j)$$ to be the number
of leading identical bits in the $7d$-bit packet representations for packets
$i$ and $j$. $\lcp(i,j)$ lies between $0$ and $7d-1$.

We will try to be consistent and use variables $i,j,t$ to denote different
packet indices and $k$ a vector-coordinate index inside an individual
packet.

\subsection{Our main construction: Shift, re-scale and add}

For a given packet index $i$ (with $0\le i<D/b$), we start with the
packet indices $j$ that are before the current packet index $i$, namely
$j=0,\ldots,i-1$. Each of the earlier packet indices $j=0,\ldots,i-1$ is
used in the following way: the vectors $v_j\in\mathbb F_p^d$ represent
the packets $B_j$. The quantity $\lcp(j,i)$ is the common-prefix length of
the $7d$-bit binary representations of $i$ and $j$. Define the {\em check
vector} $q_i\in\mathbb F_p^{8d}$ for the current packet $B_i$ as:
$$
  q_i=\sum_{j=0}^{i-1}\frac{\shift_{\lcp(j,i)}(v_j)}{i-j}
$$
Each summand is a vector in $\mathbb F_p^{8d}$, i.e. the same size as
the current check vector $q_i$. All additions and scalar multiplications
are over $\mathbb F_p$. Note that all operations (addition and scalar
multiplication) are to be performed coordinate-wise. Division by $i-j$ means
multiplication by its inverse in $\mathbb F_p$. All the packet indices are
smaller than $n<p$, so they remain distinct modulo $p$. The inverse exists
for all differences $i-j$ in the sum because
$$
  1\le i-j<p
$$
For the first packet, $q_0$ is the zero vector in $\mathbb F_p^{8d}$. Define
the {\em packet label} for the current packet $B_i$ as the pair:
$$
  Z_i=(B_i,q_i)\in\mathbb F_2^b\times\mathbb F_p^{8d}
$$

Both components are of finite length: the binary packet $B_i$ of length $b$
and the check vector of length at most $64d^2$ bits. We stress that this
pair is not a vector of a single field and that $q_i\in\mathbb F_p^{8d}$ is
not replaced by a single field element. The packet label has length $O(b)$
bits. Indeed, $q_i\in\mathbb F_p^{8d}$ has $8d$ entries, and each field
entry uses at most $8d$ bits. Thus the check vector uses at most $64d^2$
bits, which is less than $128b$ because $d^2<2b$. Including the $b$ bits of
the current packet, this yields fewer than $129b$ bits in total. Represent
every field entry by the binary expansion of its integer representative
in $\{0,\ldots,p-1\}$, using a fixed number of bits. Concatenate these
representations with $B_i$. We pad this to exactly $256b$ bits. For the
remainder of the discussion, when applying the block code, we refer to $Z_i$
as the $256b$-bit binary word in $\mathbb F_2^{256b}$.

The check vector $q_i\in\mathbb F_p^{8d}$ for packet $B_i$ uses only
information from previous packets. Therefore, we can compute the label
$Z_i=(B_i,q_i)$ after $B_i$ has been fully created. Jumping ahead to
\ref{ecc}, we will transmit $\ECC(Z_i)$ slowly during the next packet
arrival, which will still be $O(b)$ bits during the transmission of packet
$B_{i+1}$, thus maintaining a constant-rate tree code.

\section{Constant distance on packets}
\label{packet-bound}

\subsection{Comparing two packet sequences}
\label{first-difference}

\paragraph{Input packets.}
We will compare two input sequences of a fixed horizon $D$. We do
so by comparing $T$ packets of both sequences of horizon $D$, where
$1\le T\le D/b$. At positions $0,\ldots,T-1$ two packet sequences are
$B_0,\ldots,B_{T-1}$ and $B'_0,\ldots,B'_{T-1}$. For $j\in\{0,\ldots,T-1\}$,
the packets at index $j$ are $B_j,B'_j\in\mathbb F_2^b$. Call $j$ a
\emph{differing input-packet index} when
$$
  B_j\ne B'_j
$$
We always compare packets at the same position. For the purposes of
analysis, assume that the two sequences of packets are completely fixed.
We assume that there is at least one differing input-packet index. Let $a$
be the smallest such different packet $j\in\{0,\ldots,T-1\}$ with $B_j\ne
B'_j$. Define  $S$ to be the set of positions of different input packets:
$$
  S=\{j\in\{a,\ldots,T-1\}:B_j\ne B'_j\}
$$

\paragraph{Defining packet labels.}
For each packet position $t\in\{0,\ldots,T-1\}$, the packet labels at
position $t$ are:
$$
  Z_t=(B_t,q_t)
$$
$$
  Z'_t=(B'_t,q'_t)
$$
where $q_t,q'_t\in\mathbb F_p^{8d}$ are check vectors. Let $g$ count the
number of positions where $Z_t$ and $Z'_t$ differ. Since each output label
contains its current input packet as its first component, every position in
$S$ has unequal $Z$'s. Thus
$$
  1\le |S|\le g
$$
Our goal is to show that  a constant fraction of $T-a$ differ. If $a=T-1$,
this is immediate. Let's consider $a<T-1$.

\paragraph{Counting prior check vector contributions.}
Recall that for packet index $j\in\{0,\ldots,T-1\}$, vectors
$v_j,v'_j\in\mathbb F_p^d$ encode $B_j,B'_j$. They are different when
$j\in S$. $\forall j\in S$, let $r_j\in\{0,\ldots,d-1\}$ be the very first
coordinate of the vector difference where $v_j$ and $v'_j$ differ. Let
$$
  w_j=v_j[r_j]-v'_j[r_j]\in\mathbb F_p\setminus\{0\}
$$
Thus, the vector $v_j-v'_j\in\mathbb F_p^d$ is zero for all coordinates
strictly less than $r_j$ and equal to $w_j$ at coordinate $r_j$. Let's
examine packet $t$, for $a<t<T$. Define an \emph{earlier packet} as a
packet of position $j$ with $j<t$. What we want to examine is the effects of
packets at all locations $ j < t$ on the check vector $q_t$ at location $t$.
Subtracting two check vectors at location $t$ gives:
$$
  q_t-q'_t=
  \sum_{\substack{j\in S\\j<t}}
  \frac{\shift_{\lcp(j,t)}(v_j-v'_j)}{t-j}
$$
The sum only goes over $j\in S$  because all other packets have a
difference of $0$. Each part of the summation is a (shifted) vector in
$\mathbb F_p^{8d}$.
Let $$e_j(t)=r_j+\lcp(j,t)$$
Here, $e_j(t)$ is the first nonzero coordinate of packet $j$'s contribution
to $q_t-q'_t$. Its entry there is
$$
  \frac{w_j}{t-j}
$$
Indeed, shifting according to our formula moves the first nonzero entry of
$v_j-v'_j$ to its shifted coordinate, and its value is multiplied by the
inverse of the nonzero field element $t-j$.

\paragraph{Choosing the smallest coordinate that can be nonzero.}
For  $a<t<T$, consider the contributions from the prior packet to
$q_t-q'_t\in\mathbb F_p^{8d}$.  $\forall j\in S$ with $j<t$, the
contribution of (differences) of two packets at location $j$ to
(differences) of check vector at location $t$ is
$$
  \frac{\shift_{\lcp(j,t)}(v_j-v'_j)}{t-j}
$$
with $v_j,v'_j\in\mathbb F_p^d$. For packet position $j$, as we already
stated, its first nonzero coordinate is $r_j+\lcp(j,t)$. We now define
$$
  k_t=\min_{\substack{j\in S\\j<t}}
  \bigl(r_j+\lcp(j,t)\bigr)
$$
This minimum exists since $a\in S$ and $a<t$. Note that  $0\le k_t<8d$.
Define $M_t$ to be the set of  packet indices whose contributions to
$q_t-q'_t$ start at this coordinate:
$$
  M_t=\{j\in S:j<t,\ r_j+\lcp(j,t)=k_t\}
$$
Observe that  $M_t$ is a nonempty set of earlier differing input-packet
indices. All contributions to $q_t-q'_t$ are zero prior to $k_t$ coordinate.
At $k_t$,  packets indexed by $M_t$ result in nonzero entries. Thus,
$$
  q_t[k_t]-q'_t[k_t]=
  \sum_{j\in M_t}\frac{w_j}{t-j}
$$
This is the \emph{first-coordinate test} that we will use in our proof.
Observe that the sum of these contributions at the coordinate $k_t$ may
still sum up to zero and cancel. Of course, for equal output labels, this
sum is always zero.

\subsection{High-level idea for the rest of the packet distance proof}
\label{fixed-set}

First, let's recall the setup. For check vector at time $t$, for each
earlier different packet $j<t$, its contribution to $q_t-q'_t$ is at first
nonzero coordinate $e_j(t)$ and its contribution is $w_j/(t-j)$. Among all
the earlier than $t$ differing packets, the smallest such coordinate is
$k_t$. The set $M_t$ consists of the earlier differing packet numbers $j$
for which $e_j(t)=k_t$. Thus $j\in M_t$ tells us whose contribution to read,
and $k_t$ tells us where to read it in the check vector.

\paragraph{Defining the interval $J$.}
We will choose packet positions $t_0$ and $t_1$
$$
  a<t_0\le t_1<T
$$
and then define
$$
  J=\{t\in\mathbb Z:t_0\le t\le t_1\}
$$
The contiguous interval $J$ is the set of all packet positions going from
$t_0$ to $t_1$, inclusive.

Take a nonempty interval $J$ of packet times after $a$ and before $T$.
Assume (for now, without any justification) that there exists a non-empty
set $A$ of packets, all earlier than $t_0$, such that for all $t\in J$,
$M_t=A$. Then for each $t\in J$, the difference is exactly
$$
  q_t[k_t]-q'_t[k_t]=\sum_{j\in A}\frac{w_j}{t-j}
$$
since all earlier differing packets $j\notin A$ have $e_j(t)>k_t$ and
therefore contribute $0$ to coordinate $k_t$. Looking at the right-hand
side as $t$ varies, the packet numbers $j$ and their nonzero weights $w_j$
stay fixed, while $t-j$ varies. Each packet numbered $j\in A$ precedes the
first packet time in $J$. All these packet numbers and times lie between
$0$ and $T-1$, so they are all distinct modulo $p$ since $T\le n<p$. By
the reciprocal-sum ~\ref{reciprocal}, the sum is zero at at most $|A|-1$
times in $J$. Agreement of the two packet labels requires this sum to be
zero. Thus, there are at most $|A|-1$ packet times in $J$ at which the two
labels agree.

How do we choose $J$? We divide the packet times after $a$ into consecutive
intervals called \emph{phases}. In each phase, we choose one interval $J$
that covers at least one eighth of the phase and on which $M_t$ is fixed. We
will also show that each differing input packet belongs to the fixed set in
at most two phases. Thus, when we sum the bounds on equal labels across all
phases, each differing input packet is counted at most twice. This gives at
most $2|S|$ equal labels on the chosen intervals, where $S$ is the set of
differing input packet numbers.

\subsection{A Place of No Return}
\label{no-return}
We show that once a packet leaves the minimum set, it cannot return.
\begin{lemma}[No return]\label{interval}
For each $j\in S$, the set of packet times $t$ with $j<t<T$ at which
$j\in M_t$ is a consecutive interval, possibly empty.
\end{lemma}

\begin{proof}

Compare two differing input packets, an older packet $i$ and a newer packet
$j$ to the current packet $t$, so $i<j<t<T$. By property (2) of~\ref{lcp},
the newer packet's extra shift, $\lcp(j,t)-\lcp(i,t)$, never increases. The
difference between their starting coordinates in the check vector is
$$
  e_j(t)-e_i(t)=(r_j-r_i)+\bigl(\lcp(j,t)-\lcp(i,t)\bigr)
$$
Since the first term is constant, $e_j(t)-e_i(t)$ can never increase.

Suppose for a moment that packet $j$ attains the minimum at packet time
$s$. For every older differing packet $i$, we then have $e_j(s)\le e_i(s)$.
Since $e_j(t)-e_i(t)$ never increases, $e_j(t)\le e_i(t)$ at every later
packet time $t$.

Suppose $j$ leaves the minimum set at some packet time $u>s$. There must be
a differing packet $h<u$ with a strictly smaller starting coordinate at time
$u$, so $e_h(u)<e_j(u)$. By the preceding paragraph argument, $h$ cannot
be older than $j$. Thus $j<h<u$. Apply the same comparison with $h$ as the
newer packet and $j$ as the older packet. The difference $e_h(t)-e_j(t)$ is
strictly negative at $t=u$ and never increases thereafter. Thus $h$ has a
strictly smaller starting coordinate than $j$ at every later packet time,
and $j$ cannot return to the minimum set.
\end{proof}
\subsection{Choosing an interval in each phase}
\label{two-phases}

The two input streams are compared up to packet time $T-1$. Recall that $S$
is the set of packet numbers at which the streams differ, and  $a$ is the
first such packet number.

\paragraph{Defining a phase.}
A phase is a nonempty interval of consecutive packet times. We partition
$a+1,\ldots,T-1$ into phases as follows: the first phase starts at $a+1$,
if it exists. Whenever a phase starts, say at time $u$, write down the
packet numbers in $M_u$ and store this list. At each subsequent time $t$,
compare the original list $M_u$ with the current set $M_t$. As long as at
least one packet from $M_u$ appears to $M_t$, continue $t$ in the phase that
is being constructed. The next phase starts at a later (next) packet time
$v<T$ for which
$$
  M_u\cap M_v=\varnothing
$$
The current phase ends at $v-1$, and the list recorded at the start of
the next phase is $M_v$. If there is no such $v<T$, the current phase
ends at $T-1$.

\paragraph{The packet keeps the minimum throughout the phase.}
Let $w$ be the last packet time of the phase that starts at time $u$. At
least one number from $M_w$ must be on the list written down at time $u$.
To see why, suppose $w>u$. Our rule kept $w$ in the current phase since
$M_w$ contained at least one packet number from the original set $M_u$.
If $w=u$, then $M_w=M_u\ne\varnothing$. Let $j_0$ be any packet number in
$M_u\cap M_w$. Packet $j_0$ gets the minimum at time $u$ because its number
was in the original set $M_u$. It also gets the minimum at time $w$ because
$j_0\in M_w$.  Now, by the No return~\ref{interval}, once a packet leaves
the set, it can never return. If $j_0$ left at any time between $u$ and
$w$, it could not belong to $M_w$. Thus, this same packet $j_0$ attains
the minimum at every time from $u$ through $w$. It arrived before the phase
began, since $j_0\in M_u$ implies $j_0<u$. Other packets may join or leave
$M_t$ during the phase. To apply the reciprocal-sum \ref{reciprocal}, we
will find an interval covering at least one eighth of the phase on which the
whole set $M_t$ stays fixed.

\begin{lemma}[Two phases]\label{phases}
Fix a differing input packet numbered $j\in S$. Consider all later packet
times $t$ at which packet $j$ attains the minimum, that is, $j\in M_t$.
If there are any such times, they all lie in a single phase or in two
consecutive phases.
\end{lemma}

\begin{proof}
By~\ref{interval}, if $j$ attains the minimum at two packet times, it
attains the minimum at every time between them. Thus, the phases in which
$j$ attains the minimum are consecutive. Suppose that $j$ attains the
minimum in three consecutive phases. Let $u$ be the first time of the second
phase and $v$ the first time of the third. Packet $j$ attains the minimum
at some time in the first phase and again at some time in the third. It
therefore attains the minimum at both $u$ and $v$. Since $j\in M_u$, its
number was on the list recorded at the start of the second phase. Since
$j\in M_v$, one number from that list still attains the minimum at time $v$.
The rule defining a phase would therefore keep $v$ in the second phase. This
contradicts the choice of $v$ as the start of the third phase.
\end{proof}

\begin{lemma}[Left half]\label{left-half}
Every phase $F$ contains an interval $J$ of consecutive packet times
covering at least one eighth of $F$. The set $M_t$ is the same at every
time $t\in J$.
\end{lemma}

\begin{proof}
\emph{How to choose the interval $J$.} If $|F|\le 3$, take $J$ to consist of
any single time in $F$. Otherwise, choose the power of two $h$ for which
$$
  2h\le |F|<4h
$$
Let $c$ be the first multiple of $h$ in $F$. We skip fewer than $h$ times
to reach $c$, and $F$ contains at least $2h$ times. Thus the whole block
$c,\ldots,c+h-1$ lies in $F$.

Take $J$ to be the left half of this block. Its first time is $c$, and its
last time is $m-1$, where $m=c+h/2$ is the first time in the right half.
Since $|F|<4h$, the $h/2$ times in $J$ cover more than one eighth of $F$. We
will show that $M_t=M_c$ throughout $J$.

\paragraph{Which packets from $M_c$ stay at the minimum?}

By the prefix argument of property (3) of~\ref{lcp}, for every value
$j < c$, $j\in S$, the value $\lcp(j, t)$ is constant throughout the block
$[c, c + h)$, and so the coordinate $e_j(t)$ is also constant on this
block. In fact, all addresses in this block share the same block prefix, and
therefore the first bit where $j$ differs from that prefix is fixed. Recall
that one packet $j_0$ arrived before the phase began and attains the minimum
throughout the phase. Since $j_0<c$, its coordinate stays fixed on the
block. At time $c$ this coordinate is $k_c$, so the minimum stays equal to
$k_c$ throughout the block. Among the differing packets numbered before $c$,
exactly those in $M_c$ therefore continue to attain the minimum.

\paragraph{Can a packet arriving during $J$ also attain the minimum?}
Fix a packet time $t\in J$ and consider any differing packet numbered $j$
with $c\le j<t$. By property (4) of~\ref{lcp}, $\lcp(j,t) > \lcp(j, m)$
and so
$$e_j(t) > e_j(m)$$
Since $j<m$, packet $j$ is one of the packets being compared at time
$m$. The minimum at time $m$ is $k_c$, so $e_j(m)\ge k_c$. Therefore
$e_j(t)>k_c$, and packet $j$ cannot attain the minimum at time $t$.

Thus at every time $t\in J$, exactly the packets in $M_c$ attain the
minimum.
\end{proof}

\subsection{Packet Hamming Distance}
\label{count-packets}
Recall that the first differing input packet has number $a$.
\begin{lemma}[Packet distance]\label{packet-distance}
Among the $T-a$ packets at times from $a$ through $T-1$, at least a fraction
$\alpha=1/24$ have unequal labels.
\end{lemma}

\begin{proof}
Select packet time $a$. From each later phase select one interval as
in~\ref{left-half}. Let $N$ be the number of selected packet times. We take
the first time and at least one eighth of each phase, so
$$
  N\ge 1+\frac{T-a-1}{8}\ge \frac{T-a}{8}
$$

We count the selected packet times at which the two streams have equal
labels. The labels at time $a$ differ in their input packets. On a selected
interval $J$, let $A_J$ be the fixed set of packet numbers attaining
the minimum. \ref{fixed-set} bounds the number of agreements on $J$ by
$|A_J|-1$. When we sum $|A_J|$ over the selected intervals, we count
each differing input packet at most twice, since we selected one interval
per phase and each such packet attains the minimum in at most two phases
by~\ref{phases}. Thus the two labels agree at at most $2|S|$ selected packet
times, where $S$ is the set of differing input packet numbers.

Let $g$ denote the number of packet times from $a$ through $T-1$ at which
the two labels differ. Since each label contains its current input packet,
the labels differ at every time in $S$. Thus $|S|\le g$. Among the selected
packet times, at most $2g$ have equal labels and at most $g$ have unequal
labels. Together they give
$$
  N\le 3g
$$
Combining the two bounds on $N$ gives
$$
  g\ge \frac{T-a}{24}
$$
\end{proof}

\section{From packet codes to binary tree codes}

\subsection{Standard smearing trick: one output symbol per input bit}
\label{ecc}

We now use the (by now standard) smearing trick of Evans, Klugerman,
and Schulman~\cite{EvansEtAl1994}. First, we note that we know the label
$Z_i$ when packet $i$ ends. We can then send its encoded pieces during the
following packet $i+1$, one piece per arriving bit. We use a block code to
ensure that two different packet labels give different pieces at a constant
fraction of the bit times in the following packet.

\begin{lemma}[Spielman Block code]\label{block}
There are constants $K\ge 1$ and $\epsilon>0$ such that, for every positive
integer $b$, there is a deterministic encoder
$$
  \ECC_b:\mathbb F_2^{256b}\longrightarrow(\mathbb F_2^K)^b
$$
for which the codewords of two distinct messages differ in at least
$\epsilon b$ of their $b$ pieces. Constructing the encoder takes $b^{O(1)}$
time. Encoding one message takes $O(b)$ word operations.
\end{lemma}

\begin{proof}
Use Spielman's family of constant-rate binary codes with constant relative
distance and linear-time encoding \cite[Theorem~19]{Spielman96}. Pad the
message with zeros to the next supported message length, which is $O(b)$.
The family is constructible in polynomial time. Each codeword has $O(b)$
bits, and the codewords of two distinct messages differ in $\Omega(b)$
bit positions.

Choose a constant integer $K$ large enough to pad each codeword with zeros
to $Kb$ bits. Split these bits into $b$ pieces of $K$ bits each. Since a
piece holds only $K$ bits, the differing bit positions occupy $\Omega(b)$
pieces.
\end{proof}

Write $C_{i,s}$ for piece number $s$ of $\ECC_b(Z_i)$. Packets are numbered
from $0$. Within a packet, bits and codeword pieces are numbered from $1$
through $b$. When bit $s$ of packet $i$ arrives, the bit time is $t=ib+s$,
and the output symbol is
$$
  (\Local_t,C_{i-1,s})
$$
The first component is the local label from \ref{second-encoder} with
$r=d$. For $i\ge 1$, the second component is piece $s$ of the preceding
packet's codeword. For $i=0$, define $C_{-1,s}=0^K$. Both components
are available when the bit arrives. The same alphabet $\Gamma=\mathbb
F_2^{49}\times\mathbb F_2^K$ works for every packet length.

\subsection{Putting it together: distance at a finite horizon
  \texorpdfstring{$D$}{D}}
\label{finite}

We denote $\Fin_D(x_1,\ldots,x_t)$ the output symbol at bit time $t$ for
$1\le t\le D$. The local labels protect short-run  comparisons.  For longer
comparisons, enough whole packets allow us to transmit the codewords of
earlier packet labels.

\begin{lemma}[Finite distance]\label{finite-lemma}
The online encoder $\Fin_D$ has relative distance at least
$$
  \delta=\min\{1/16,\alpha\epsilon/2\}
$$
where $\alpha$ is the packet distance from~\ref{packet-distance} and
$\epsilon$ is the block-code distance from~\ref{block}.
\end{lemma}

\begin{proof}
Compare two streams from their first differing bit through the last bit
under consideration. Let $L$ be the number of bit times in this interval.
If $L\le 4b$, consider the first $\min\{L,b\}$ bits of this interval. This
covers at least $L/4$ bit times. It also fits within the local encoder's
guarantee since $b\le d^2$. By~\ref{local}, at least a quarter of the local
labels there differ. This gives at least $L/16$ unequal output symbols.
Suppose $L>4b$. Let $i$ be the number of the packet containing the first
differing bit. Skip the rest of packet $i$, including that bit. At the
other end, drop any incomplete packet. We lose at most $b$ bit times at
the start and fewer than $b$ at the end. Thus the $m$ whole packets left
cover more than $L/2$ bit times. These $m$ whole packets are numbered
$i+1,\ldots,i+m$. Each transmits the ECC codeword for the preceding packet's
label, so together they transmit codewords for packets $i,\ldots,i+m-1$.
~\ref{packet-distance} guarantees at least $\alpha m$ unequal packet
labels among these $m$ labels. By~\ref{block}, each unequal label produces
at least $\epsilon b$ unequal symbols in the following packet. Thus the
number of unequal symbols is at least $\alpha\epsilon mb$, which exceeds
$\alpha\epsilon L/2$.
\end{proof}

\paragraph{Setting up parameter sizes.}
The following two observations follow from the parameter settings in
~\ref{overview}. First, the horizon of $D$ input bits contains a whole
number of packets. Indeed, both $b$ and $D$ are powers of two, so it
suffices to check $b\le D$. For $d=1$ we have $b=1$. For $d\ge 2$,
the preceding candidate $d-1$ had capacity less than $D$. It allowed
$2^{7(d-1)}$ packets of at least one bit each. Thus
$$
  D>2^{7(d-1)}\ge d^2\ge b
$$
Observe that  $nb$ is within a constant factor of $D$. Increasing $d$ by one
multiplies $n$ by $128$ and $b$ by at most 4. We stop when $nb\ge D$, and
the initial capacity is $128$. Since $b=\Theta(d^2)$ and $n=2^{7d}$, we get
$$
  nd^2=\Theta(D)
$$
Taking logs yields $d=\Theta(\log(D+2))$.

\subsection{Two overlapping horizons for \texorpdfstring{$O(1)$}{O(1)}
  infinite tree code}
\label{infinite}
It is shown in \cite[Lemma A.1]{BenYaacovEtAl2022} that, given a collection
of finite tree codes $\mathsf{TC}_{2^i}$ with distance $\delta$ and
alphabet $\Gamma$, for every $i \in \mathbb{N}$, there exists an infinite
tree code with distance $\delta/2$ and alphabet $\Gamma^3$. Based on the
particulars of our construction, we give a modified finite-to-infinite
construction here.

To encode a stream of any length, use the finite encoders in pairs, similar
to~\cite{BenYaacovEtAl2022}. Choose a power of two $\rho\ge 2/\delta$,
where $\delta$ is from \ref{finite-lemma}. For any given bit time $t$, let
$D(t)$ be the smallest power $\rho^k$, $k\in \mathbb{Z}_{>0}$, that exceeds
$t$. Output
$$
  E(x_1,\ldots,x_t)=
  \bigl(\Fin_{D(t)}(x_1,\ldots,x_t),\Fin_{\rho D(t)}(x_1,\ldots,x_t)\bigr)
$$
Note that each of the finite encoders uses the entire input prefix. When the
bit time reaches the smaller horizon $D$, the pair of horizons changes from
$(D,\rho D)$ to $(\rho D,\rho^2D)$. Thus, the encoder with horizon $\rho D$
becomes the first component. This overlap keeps one encoder visible across
the switch. The alphabet becomes $\Gamma^2$.

\begin{theorem}
The infinite tree code given above has a distance of at least $\delta/2$.
\end{theorem}
\begin{proof}
Let $a$ be the point of first divergence between two input streams, and
let $D$ be the smallest power $\rho^k$ that exceeds $a$. Fix $T>a$. We
compare outputs at $T-a$ bit times $a,a+1,\ldots,T-1$. Our aim is to show
that two outputs differ by at least $\delta(T-a)/2$ positions in this
interval. Consider the interval $[a, T)$, and let $k$ be the smallest
integer such that $\rho^k D$ exceeds $T$. If $k \leq 1$, use the tree
code $\Fin_{\rho D}$ on the entire interval $[a, T)$, giving a distance of
$\delta$. For $k > 1$, the tree code $\Fin_{\rho^k D}$ will be used on the
interval $[\rho^{k-2}D, T).$ The tree code $\Fin_{\rho^k D}$ has distance
at least $\delta$ on the interval $[a, T)$, but the points of difference may
be concentrated on the interval $[a, \rho^{k-2}D)$, which is missing from
our infinite code. This interval is small enough, however, so we can bound
the distance by
$$
  \frac{\delta (T-a) -(\rho^{k-2} D -a)}{T - a} > \delta - \frac{\rho^{k-2}
  D}{T} \geq \delta - \frac{\rho^{k-2}}{\rho^{k-1}} \geq \delta -
  \frac{1}{\rho}
  \geq\frac{\delta}{2}
$$
as desired.
\end{proof}

\section{Computational Results}

\subsection{Computing a single label more efficiently}
\label{computation}

\begin{lemma}
All check vectors at depths up to $t$ can be computed in
$O(t\cdot\mathrm{polylog}(t))=\widetilde{O}(t)$ bit operations.
\end{lemma}

\begin{proof}
Assume $0\le j_0\le j_1<t_0\le t_1<n<p$. Once the reciprocals have been
specified, the sums
$$
  \sum_{j=j_0}^{j_1}\frac{w_j}{i-j}
$$
for $t_0\le i\le t_1$, it can be computed in a total time of
$\widetilde{O}((j_1-j_0+1)+(t_1-t_0+1))$, counting field operations. Recall
that the field arithmetic costs $\widetilde{O}(\log p)$ operations on bits.
To compute them, multiply the formal polynomials
$$
  \left(\sum_{j=j_0}^{j_1}w_jX^j\right)
  \left(\sum_{s=t_0-j_1}^{t_1-j_0}\frac{X^s}{s}\right)
$$
The coefficient of $X^i$ is equal to the desired sum at $i$, since a term
of degree $i$ has $s=i-j$. We factor out the lowest power of $X$ in each
polynomial before multiplying to obtain the stated cost. For a horizon $H$,
the block calculation below covers every pair of packets $j<i$ in $O(H)$
word operations in total, including setup. The horizons through input depth
$t$ increase geometrically, and the largest is $O(t)$, so summing their
costs proves the lemma.
\end{proof}

For $j<i-1$, choose the largest power of two $h$ such that at least one full
length-$h$ block lies between the aligned blocks containing $j$ and $i$ at
that scale. Because the parents of the two blocks are adjacent, the number
of target blocks per source block is at most two at that scale. For each
pair of packets within a block pair (one packet from each block), the same
coordinate shift is used. We compute these weighted sums by convolution,
using an FFT on integer representatives and reducing the resulting
sums modulo $p$. As each packet finishes arriving, the encoder adds its
contribution directly to the next packet's check vector.

Pack the $d$ coordinates of a packet vector $v_j$ into one word using
$\lceil\log_2(np^2+1)\rceil$ bits per coordinate to prevent carries between
coordinates. The terms to be convolved are integers, so each coefficient
$c_k$ of the convolution is an integer. Thus, we only need to ensure that
the computed approximation is within $1/2$ of $c_k$. The result, rounded to
the nearest integer, will then be the correct coefficient. Set $U$ to be the
maximum absolute value of a coefficient in either input polynomial. With $P$
fractional bits, the FFT error is at most $2^{-P}(2h)^{O(1)}(U+1)^2$. Since
$h\le 2^{7d}$ and $U=2^{O(d^2)}$, taking $P=Cd^2$ makes this error less
than $1/2$ for a sufficiently large constant $C$. Each FFT stage consists
of updates called \emph{butterflies}. For each butterfly with inputs $u$
and $v$, we compute the values $u+\omega v$ and $u-\omega v$, where $\omega$
is a root of unity. The magnitude of $\omega$ is $1$. For input magnitudes
of at most $V$, the exact values after $s$ stages have a magnitude of at
most $2^sV$.

Let $E_s$ be the largest absolute difference between a computed value and
its exact value after stage $s$. With $P$ fractional bits, the rounded
root has a magnitude of at most two. The errors in $u$ and $v$ therefore
contribute at most $E_s+2E_s$ to the error in each butterfly output.
Rounding the root and the arithmetic in the butterfly contributes another
$O(2^{-P}(2^sV+1))$. Thus,
$$
  E_{s+1}\le 3E_s+O\bigl(2^{-P}(2^sV+1)\bigr)
$$
Iterate over the $\log_2(4h)$ stages of each length-$4h$ transform, using
$4hU$ to bound the magnitude of each exact forward output in the pointwise
products.

The FFT also needs a table of roots of unity. For each root, we compute
its two coordinates, the cosine and sine of an angle in $[0,2\pi]$. On
this interval, $O(P)$ terms of each series suffice to make the truncation
error at most $2^{-P-2}$. As we compute those terms, small rounding errors
can accumulate. We therefore keep a few extra bits after the binary point
during the calculation. These extra bits are called \emph{guard bits}.
Each extra bit halves the bound on the error due to rounding. If we keep
$P+g$ fractional bits (where $g$ are guard bits) after the binary point,
the accumulated rounding error is $O(P\,2^{-(P+g)})$. We can thus choose
$g=\lceil\log_2(P+2)\rceil+c$ for a sufficiently large constant $c$ to
keep this error below $2^{-P-2}$. Each coordinate is then rounded to $P$
fractional bits, giving a final error of at most $2^{-P}$. Each series term
takes $O(1)$ word operations to compute from the preceding term, so each
root costs $O(P)$ word operations, and all $O(n)$ roots cost $O(nP)$ word
operations. The FFT costs $O(h\log(2h))$ word operations, and $O(d^2)$ bits
of precision are sufficient for exact rounding. Unpacking and shifting costs
$O(hd)$ word operations. Spreading this work over the first $h$ packets in
the gap costs $O(d+\log(2h))$ per packet at this scale. Summing over the
$O(d)$ scales gives $O(d^2)=O(b)$ work per packet.

At horizon $H$, it takes $O(nd^2)=O(H)$ word operations to run the
short-code search and the prime sieve, construct the lookup table for
the $2^d$ local histories, the reciprocal and FFT tables, and set up the
block code. Local lookup cost is $O(1)$ per bit, and block encoding costs
$O(b)$ per packet. A horizon of length $H$ contains $H/b$ packets, and both
horizons used at time $t$ have a length of $O(t)$. In total, the label at
time $t$ can be computed in $O(t)$ word operations or in $O(t\log^4(t+2))$
(and hence $O(t^2)$) bit operations using schoolbook arithmetic.

\subsection{Computing successive labels online using pipelining}
\label{pipelining}

Starting the new encoder from scratch when the horizon changes would require
processing the entire input prefix at the switch. To meet the per-bit bound
in \ref{main}, we pre-compute the next horizons slowly. This is what we call
{\em pipelining.}

The initial horizons cost $O(1)$. Build each subsequent horizon $H$ before
its first use at $A=H/\rho^2$. Spread its $O(H)$ setup over $[A/4,A/2)$.
Since $\rho$ is fixed, this costs $O(1)$ word operations per arrival. During
$[A/2,A)$, replay up to two buffered bits per arrival in order. At time $A$,
the new encoder has processed the first $A-1$ bits and can process bit $A$
as it arrives.

Only constantly many horizons overlap, and each processes at most two bits
per arrival, giving $O(\log^2(t+2))$ worst-case word operations per arrival.
The horizons started by time $t$ grow geometrically, and each has size
$H\le 4\rho^2t$, since its setup starts at $H/(4\rho^2)$. Summing the $O(H)$
work over these horizons gives $O(t)$ total work.

\section*{Acknowledgments}
We thank Brett Hemenway Falk for discussions in 2024 on the tree code
problem. AI was used to search the literature and check for errors.

\bibliographystyle{alpha}
\bibliography{refs}

\end{document}